\documentclass[twocolumn,aps,prl,epsfig,showpacs,superscriptaddress]{revtex4}

\usepackage{graphicx}

\begin{document}

\title{Co-doped (La,Sr)TiO$_{3-\delta}$ : a high-Curie temperature diluted magnetic system with large spin-polarization}

\author{G. Herranz}
\address{Unit\'e Mixte de Physique CNRS/Thales, Route d\'epartementale 128, 91767 Palaiseau Cedex, France}

\author{R. Ranchal}
\address{Departamento de F\'{i}sica de Materiales, Facultad de Ciencias F\'{i}sicas (UCM), Avenida Complutense, s/n
28040-Madrid, Spain }
\author{M. Bibes}
\address{Institut
d'Electronique Fondamentale, Universit\'e Paris-Sud, 91405 Orsay, France}
\author{H. Jaffr\`es}
\address{Unit\'e Mixte de Physique CNRS/Thales, Route d\'epartementale 128, 91767 Palaiseau Cedex, France}
\author{E. Jacquet}
\address{Unit\'e Mixte de Physique CNRS/Thales, Route d\'epartementale 128, 91767 Palaiseau Cedex, France}
\author{J.-L. Maurice}
\address{Unit\'e Mixte de Physique CNRS/Thales, Route d\'epartementale 128, 91767 Palaiseau Cedex, France}
\author{K. Bouzehouane}
\address{Unit\'e Mixte de Physique CNRS/Thales, Route d\'epartementale 128, 91767 Palaiseau Cedex, France}
\author{F. Wyczisk}
\address{Thales Research and Technology France, Route d\'epartementale 128, 91767 Palaiseau Cedex, France}
\author{E.Tafra}
\address{Department of Physics, Faculty of Sciences, University of Zagreb, Bijenicka 32 - P.O.B. 331 HR-10002 Zagreb,
Croatia}
\author{M. Basletic}
\address{Department of Physics, Faculty of Sciences, University of Zagreb, Bijenicka 32 - P.O.B. 331 HR-10002 Zagreb,
Croatia}
\author{A. Hamzic}
\address{Department of Physics, Faculty of Sciences, University of Zagreb, Bijenicka 32 - P.O.B. 331 HR-10002 Zagreb,
Croatia}
\author{C. Colliex}
\address{Laboratoire de Physiques des Solides, Universit\'e Paris-Sud-UMR 8502, 91405 Orsay, France}
\author{J.-P. Contour}
\address{Unit\'e Mixte de Physique CNRS/Thales, Route d\'epartementale 128, 91767 Palaiseau Cedex, France}
\author{A. Barth\'el\'emy}
\address{Unit\'e Mixte de Physique CNRS/Thales, Route d\'epartementale 128, 91767 Palaiseau Cedex, France}
\author{A. Fert}
\address{Unit\'e Mixte de Physique CNRS/Thales, Route d\'epartementale 128, 91767 Palaiseau Cedex, France}

\date{\today}

\begin{abstract}

\vspace{0.5cm}

We report on tunneling magnetoresistance (TMR) experiments that demonstrate the existence of a significant spin
polarization in Co-doped (La,Sr)TiO$_{3-\delta}$ (Co-LSTO), a ferromagnetic diluted magnetic oxide system (DMOS)
with high Curie temperature. These TMR experiments have been performed on magnetic tunnel junctions associating
Co-LSTO and Co electrodes. Extensive structural analysis of Co-LSTO combining high-resolution transmission
electron microscopy and Auger electron spectroscopy excluded the presence of Co clusters in the Co-LSTO layer
and thus, the measured ferromagnetism and high spin polarization are intrinsic properties of this DMOS. Our
results argue for the DMOS approach with complex oxide materials in spintronics.

\end{abstract}

\maketitle

The research of ferromagnetic semiconductors is an important challenge in nowadays spintronics. Since
ferromagnetic properties have been found in Mn-GaAs \cite{ohno96}, several types of other diluted magnetic
semiconductors have been investigated. The discovery of room temperature ferromagnetism in Co-doped anatase
TiO$_2$ \cite{matsumoto2001} has triggered an intense research on other diluted magnetic oxide systems (DMOS).
Some recent experiments have also suggested the existence of ferromagnetism in
La$_{0.5}$Sr$_{0.5}$TiO$_{3-\delta}$ doped with Co \cite{zhao2003,qiao2004} i.e. a DMOS in which the host oxide
is not a semiconductor but a strongly correlated metal \cite{tokura93}. Several models of ferromagnetism have
been proposed for these DMOS, including models based on a new type of exchange interactions via shallow donors
(associated with oxygen vacancies) \cite{coey2005}.

Application of DMOS to spintronics requires not only ferromagnetism but also spin polarization of the carriers.
This spin polarization can be studied by several techniques. For instance, X-ray magnetic circular dichroism is
one of the best tools to detect the coupling between delocalized sp-carriers and localized d-electrons and was
successfully applied in transition metal doped ZnO \cite{ando2001}. On the other hand, it has been argued that
anomalous Hall effect (AHE) should be observed in these compounds. Nevertheless, the concomitant observation in
Co-doped TiO$_2$ of parasitic superparamagnetic metallic Co clusters and AHE has cast doubts on the usefulness
of AHE to unambiguously demonstrate intrinsic spin-polarization in DMOS \cite{shinde2004}. A more direct way is
provided by tunneling magnetoresistance (TMR) measurements that can probe directly the spin polarization
involved in spintronics experiments. This is the way we have taken to investigate the ferromagnetism and the
spin polarization in the (La,Sr)Ti$_{1-x}$Co$_x$O$_{3-\delta}$ (Co-LSTO, x$\simeq$0.01) DMOS.

In this Letter we thus report on the growth of Co-LSTO films, their magnetic and structural characterizations,
on the fabrication of tunnel junctions composed of Co-LSTO and Co electrodes separated by a LaAlO$_3$ (LAO)
barrier and on the TMR experiments performed on these junctions and their interpretation.

We have deposited Co-LSTO epitaxial thin films on SrTiO$_3$ (001) substrates by pulsed laser deposition (PLD)
\cite{ranchal2005} from a target with composition La$_{0.37}$Sr$_{0.63}$Ti$_{0.98}$Co$_{0.02}$O$_3$ (i.e.
x$\simeq$0.02). The heterostructures elaborated for the definition of MTJs were  fabricated by first growing
Co-LSTO(150 nm)/LAO(1.6 nm) bilayers at P$_{O_2}$ = 6$\times$10$^{-7}$ mbar by PLD and subsequently
sputter-depositing ex-situ a Co(12 nm)/CoO(3nm)/Au(15 nm) stack. The CoO was intended to magnetically pin the
top Co-electrode by exchange bias. The heterostructures were then patterned into MTJs with size ranging from 6
to 128 $\mu$m$^2$ by an advanced photolithography process \cite{bowen2003}.

\begin{figure}
\includegraphics[width=0.9\columnwidth]{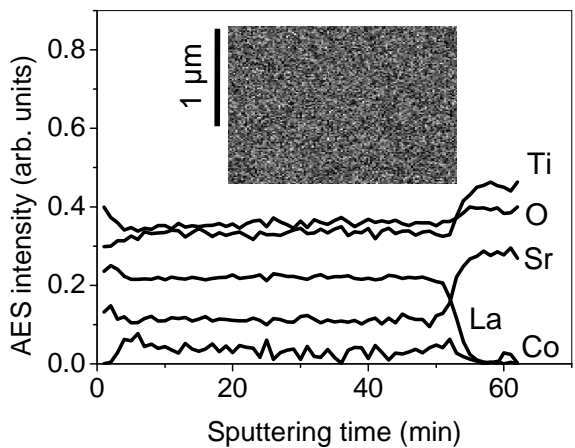}
\caption{Auger intensity for several elements, measured as a function of the sample etching time, for a Co-LSTO
film of thickness t = 130 nm grown at P$_{O_2}$ = 6$\times$10$^{-7}$ mbar. Inset : AES mapping of Co in the same
film.} \label{auger}
\end{figure}

We have investigated the structural properties and the homogeneity of the chemical composition of various
Co-LSTO films by combining several advanced characterization tools. Auger electron spectroscopy (AES) mappings
were performed over regions of size up to 10$\times$10 $\mu$m$^2$ to detect La, Sr, Ti, O and Co elements. The
resulting elemental mappings were homogeneous and did not show any contrast (see for instance a mapping for Co
in the inset of Fig. \ref{auger}). The AES signals were also measured in real-time while sputtering the films
with Ar ions accelerated at 2 keV, see figure \ref{auger}. The AES signals remain roughly constant, indicating a
homogeneous distribution of the different elements, until a sputtering time of about 50 min when the La and Co
levels decay steadily to zero, while the Sr and Ti levels increase and then stay constant. This signals the
interface between the Co-LSTO film and the STO substrate. The level of the Co was estimated to be around x=0.01,
which is somehow lower than the nominal x = 0.02.

\begin{figure}
\includegraphics[width=0.9\columnwidth]{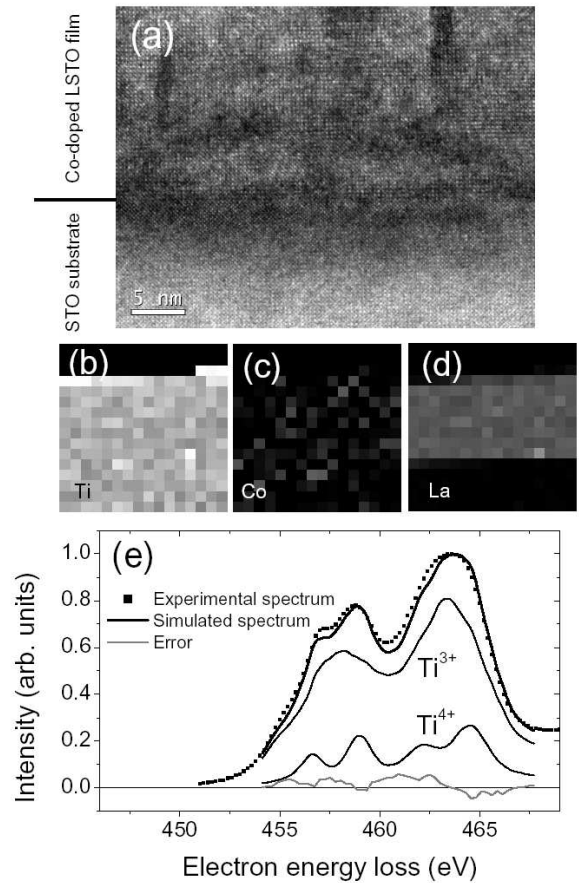}
\caption{(a) HRTEM image of a LSTO//STO system (t = 100 nm, P$_{O_2}$ = 6$\times$10$^{-7}$ mbar) in cross
section in [100] zone. (b-d) EELS element-selective images. Grey levels are proportional to the
[element]/[oxygen] ratio. Black is zero; white is 50\% in Ti and La images, 1\% in Co image. The pixel size is
12 nm. (e) Experimental and simulated EELS spectra at the Ti-L$_{2,3}$ edge, together with reference spectra for
Ti$^{3+}$ and Ti$^{4+}$ (adapted from reference \onlinecite{ohtomo2002}).} \label{tem}
\end{figure}

We have also investigated the microstructure of Co-LSTO thin films using transmission electron microscopy (TEM)
and high-resolution TEM (HRTEM) experiments combined with electron energy loss spectroscopy (EELS) measurements.
No parasitic phase (e.g. Co clusters) was detected by TEM or HRTEM. In Fig. \ref{tem}a we show a HRTEM image of
the Co-LSTO/STO interface in the [100] zone evidencing a good epitaxial quality. The absence of Co clusters was
further confirmed by low-magnification EELS elemental mappings in cross section (see in Fig. \ref{tem}b-d). The
signal in the Co image is pure noise in the used recording conditions (0.5 s pixel$^{-1}$).

Finally we have also analyzed the fine structure of the Ti-edge in the EELS spectra, which is characteristic of
its valence. We estimate the 3+/4+ ratio in our samples by considering the Ti-edges we have recorded (Fig.
\ref{tem}e), as a linear combination of the edges of pure Tî$^{3+}$ and Ti$^{4+}$ recorded on standards with
approximately the same energy resolution (0.8 eV) . We find a 3+/4+ ratio of $\sim$3:1, corresponding to an
average valence of $\sim$3.25+ for the Ti ions. If we assume the La/Sr ratio to be $\sim$0.58 as in the target,
the Ti valence should be $\sim$3.63+. This difference can be explained by the presence of oxygen vacancies
promoted by the low growth pressure, creating electrons that populate the Ti 3d band. The oxygen deficiency
necessary to account for a valence of 3.25+ is $\delta\simeq$0.19 that is close to the value found by Muller et
al, in SrTiO$_{3-\delta}$ films grown in similar conditions \cite{muller2004}.

\begin{figure}
\includegraphics[width=0.9\columnwidth]{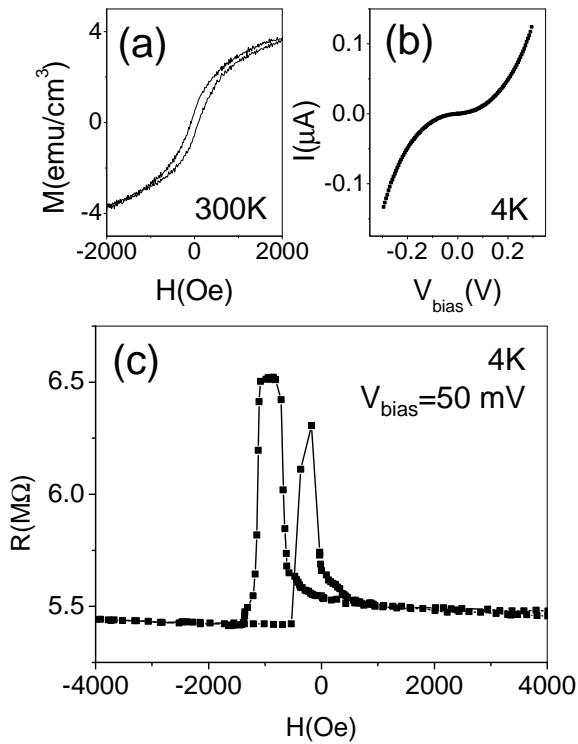}
\caption{(a) Magnetization hysteresis cycle measured with an alternating gradient force magnetometer (AGFM) for
a 130 nm Co-LSTO film grown at P$_{O_2}$=6$\times$10$^{-7}$ mbar, with the magnetic field applied in plane. (b)
I-V curve of a Co-LSTO/LAO/Co tunnel junction measured at 4K and in a field H = 6000 Oe. (c) Field dependence of
the resistance of a Co-LSTO/LAO/Co magnetic tunnel junction (size 64 $\mu$m$^2$), at 4K and bias-voltage of 50
mV.} \label{tmr}
\end{figure}

To summarize this structural part, we can conclude from the HRTEM pictures and the elemental mappings obtained
by AES and EELS that the films are of very good structural quality, with no extended defects and, more
importantly, that the Co distribution is homogeneous in the film. No Co-rich regions larger than $\sim$10 nm are
present.

In Fig. \ref{tmr}a, we show a typical ferromagnetic hysteresis cycle measured at room temperature for a 130 nm
Co-LSTO film grown at P$_{O_2}$=6$\times$10$^{-7}$ mbar. The saturation magnetization (M$_S$) for this film is
about 4.5 emu.cm$^{-3}$ and was found to increase when P$_{O_2}$ decreased from P$_{O_2}$=5$\times$10$^{-4}$ to
6$\times$10$^{-7}$ mbar. This suggests some role of oxygen vacancies on magnetism. From the Co concentration
measured by AES, we can calculate that the magnetization of the low-P$_{O_2}$ films corresponds to 3-4
$\mu_B$/Co, which is fairly larger than the moment of metallic Co (1.7 $\mu_B$). This may be an indication of
magnetic interactions between Co ions mediated by carriers at the 3d-Ti conduction band. This picture is
supported by some preliminary nuclear magnetic resonance (NMR) measurements on our Co-LSTO films showing the
presence of ferromagnetic Ti, as indicated by observation of $^{47}$Ti and $^{49}$Ti NMR signals corresponding
to a hyperfine field of 12T at 4.2 K \cite{jedryka2005}.

Figure \ref{tmr}c shows a TMR curve recorded at 4 K in one of our Co-LSTO/LAO/Co/CoO/Au MTJs (size: 64
$\mu$m$^2$). The curve was obtained at V$_{bias}$ = 50 mV after field cooling in 6 kOe. The R(H) curves are
shifted respect to the H = 0 axis, which is due to the exchange bias induced by the CoO layer on the Co
electrode and, for the Co-LSTO electrode, to the dipolar field generated by the Co layer in the mesa structure
of the junction. Note that a plateau is clearly observed in the antiparallel state, which suggests that the
magnetization of the Co-LSTO layer is almost saturated in this field range under the action of the applied field
and the dipolar field induced by the Co layer. The I(V) curves (see Fig. \ref{tmr}b) are non-linear, indicative
of tunnel transport. Our first conclusion is that large TMR effects (up to $\sim$19\%) are observed and
demonstrate the spin polarization of the carriers in the Co-LSTO electrode. The existence of such
spin-polarization as well as a robust magnetism for a Co concentration of $\sim$1 at\% allows excluding
short-range interactions such as conventional super-exchange or double-exchange to be at the origin of
ferromagnetism in this compound. We discuss below the quantitative determination of the spin polarization.

\begin{figure}[h!]
\includegraphics[width=0.9\columnwidth]{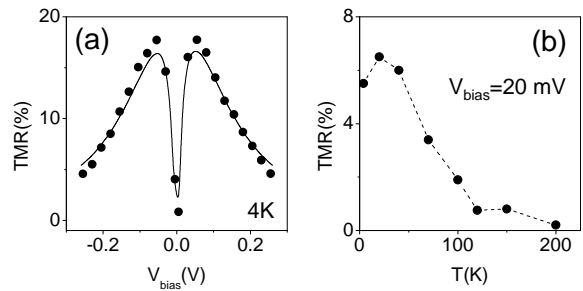}
\caption{(a) TMR bias dependence of a 64 $\mu$m$^2$ Co-LSTO/LAO/Co MTJ. The solid line correspond to the fit
obtained taking into account a Lorentzian damping function with a characteristic value V$_{1/2}$=150 mV and the
following parameter values: $P_L$= -0.3, $P_R$= -0.78, $\Gamma_L$ =
($\Gamma_{L\uparrow}$-$\Gamma_{L\downarrow}$)/2 = 0.88 meV, $\Gamma_R$ =
($\Gamma_{R\uparrow}$-$\Gamma_{R\downarrow}$)/2 = 5.7 meV, $\varepsilon_C$=1.65 meV. Optimal fittings were
obtained by assuming a narrow energy band ($W$ = 0), corresponding to the case of impurities occupying the same
energy level (i. e., the mean spatial separation of impurities might be too large to induce any overlapping and
hybridization between two neighboring defects). (b) Temperature dependence of a Co-LSTO/LAO/Co junction (size :
12 $\mu$m$^2$) at bias V = 20 mV. } \label{tmr-v-t}
\end{figure}

A striking result of our tunneling experiments is the bias dependence of the TMR, see figure \ref{tmr-v-t}a.
This type of bias dependence is usually observed when there is a significant contribution of impurity-assisted
(or defect-assisted) tunneling. A clear interpretation of this bias dependence has been given by Tsymbal et al.
for Ni/NiO/Co nanojunctions in which the small size (0.01 $\mu$m$^2$) of the junctions allowed these authors to
assume resonant tunneling via a single localized level (a single level in a given junction) \cite{tsymbal2003}.
Here we use the extension of the model worked out by Garcia et al. \cite{garcia2005} to describe the situation
with still single-impurity-assisted tunneling but localized levels distributed in a band. In our junctions, the
localized levels could be due to O vacancies promoted by the low oxygen pressure growth conditions. We assume a
distribution of localized levels within a band of width $W$, centered at a mean energy $\varepsilon_C$ from the
Fermi energy level $E_F$ of the electrodes and having a Lorentzian energy distribution. The conductance per
impurity can be expressed as:

\begin{equation}
G(\sigma,\sigma')=\frac{4e^2}{h}\frac{\Gamma_{L\sigma}\Gamma_{R\sigma'}}{\Gamma_{L\sigma}+\Gamma_{R\sigma'}}
\frac{\Gamma_{L\sigma}+\Gamma_{R\sigma'}+W}{(\varepsilon-\varepsilon_C)^2+(\Gamma_{L\sigma}+\Gamma_{R\sigma'}+W)^2}
\end{equation}

\noindent where $\Gamma_{L\sigma}/\hbar$, $\Gamma_{R\sigma'}/\hbar$ are the leak rates of an electron from the
localized state into the left (L) or right (R) electrode, respectively, and $\sigma$, $\sigma'$ refer to the
spin state ($\uparrow,\downarrow$) of the left (right) electrode. This resonant tunneling model includes six
parameters, namely, $\Gamma_{L\sigma}$, $\Gamma_{R\sigma'}$, $W$, $\varepsilon_C$. The spin polarizations of the
leak rates, $P_L$=($\Gamma_{L\uparrow}-\Gamma_{L\downarrow}$)/($\Gamma_{L\uparrow}+\Gamma_{L\downarrow}$) and
$P_L$=($\Gamma_{R\uparrow}-\Gamma_{R\downarrow}$)/($\Gamma_{R\uparrow}+\Gamma_{R\downarrow}$) are the spin
polarizations of the left and right electrodes. All the TMR bias dependences can be fitted with Eq. (1) and with
a convenient set of parameters, which indicates that the main contribution to the tunneling comes from this
process. We show a typical example in Fig. \ref{tmr-v-t}a. The most remarkable result is that we extract almost
the same spin-polarization for each electrode from fits of the bias dependence of TMR in several junctions, i.e.
$P_L$=-0.25$\pm$0.05 and $P_R$= -0.8$\pm$0.02 (or $P_L$=+0.25$\pm$0.05 and $P_R$= +0.8$\pm$0.02), since we
cannot determine the sign of each polarization. However, preliminary TMR measurements obtained on
La$_{2/3}$Sr$_{1/3}$MnO$_3$/LAO/Co MTJs indicate a negative spin polarization of about $P\simeq$ -15 \% at the
Co/LAO interface. We can thus consider that both $P_L$ and $P_R$ are negative and that the smallest
spin-polarization (in absolute value) corresponds to that of the Co/LAO interface, that is $P_{Co/LAO}$=
-0.25$\pm$0.05. This therefore implies a spin-polarization of -0.8$\pm$0.02 for the Co-LSTO/LAO interface. This
remarkably high spin polarization found at the Co-LSTO/LAO interface undoubtedly argues for an intrinsic origin
of ferromagnetism in Co-LSTO.

The last question is the origin of these impurity states. We recall that in order to optimize the magnetic
properties of the Co-LSTO films the Co-LSTO/LAO bilayers were grown at extremely low oxygen pressures (6$\times$
10$^{-7}$ mbar). Thus, the barriers were not grown in standard conditions, and very likely contain oxygen
vacancies that create defect states within the LAO gap. In Fig. \ref{tmr-v-t}b we present the temperature
dependence of the TMR of a 12 $\mu$m$^2$ junction at V$_{bias}$ = 20 mV. The TMR decreases rapidly with T and
becomes vanishingly small above 200 K. This behavior is unexpected since Co-LSTO is ferromagnetic at 300 K (see
Fig. \ref{tmr}a) and the Curie temperature of Co is far above room temperature. There are several possible
origins for this behavior. First, we note that the pinning by exchange bias of the Co-top electrode in our
structures disappears at T $>$ 150 K, which can lead to very similar values of the coercive fields of Co and
Co-LSTO above this temperature and, consequently, to the absence of a field range with antiparallel orientation.
Another origin may be related to the onset of thermally assisted spin flips on impurity levels inside the
barrier, which may be detrimental to the spin polarization of the tunneling current as observed in
MnAs/AlAs/MnAs MTJs \cite{garcia2005}. In a model of resonant tunneling through localized states as well as in
the case of sequential processes through an energy band, TMR subsists unless the tunneling time
$\tau_n=\hbar/(\Gamma_L+\Gamma_R)$ of the particle exceeds its spin lifetime $\tau_{sf}$. The rapid decrease of
TMR with temperature might indicate the effects of some thermally-assisted spin flip mechanisms drastically
shortening the spin lifetime with temperature. Finally one must not discard the possibility of a degradation of
the spin polarization at the Co-LSTO/LAO interfaces, as observed in other ferromagnetic oxides
\cite{bibes2001e}. More experiments are foreseen in order to clarify the role of the Co-top electrode pinning or
the impurity levels in the barrier in both the bias and temperature dependence of the MR of Co-LSTO-based MTJs.

In summary, we have reported on TMR results indicating a significant spin polarization of the diluted magnetic
oxide system Co-LSTO. By HRTEM, EELS, AES and AGFM measurements, we exclude Co-segregation as the origin of the
ferromagnetism and TMR in Co-LSTO, and we conclude that the spin-polarization as well as the ferromagnetism are
intrinsic properties of this DMOS. Our results bring strong arguments for the DMOS approach with complex oxide
materials in spintronics.

We are grateful to P. Berthet, C. Pascanut and N. Dragoe for providing the Co-LSTO target. G. Herranz
acknowledges financial support from Minist\`ere de l'Education Nationale, de l'Enseignement Sup\'erieur et de la
Recherche (France) and R. Ranchal thanks Universidad Complutense de Madrid and the Spanish Project MAT
2001-3554-CO2 for partially supporting her stay at UMR137, CNRS-Thales at Orsay, France.

\vspace{0.5em}

$^{*}$ corresponding author : gervasi.herranz@gmail.com.

\vspace{0.5em}

\bibliographystyle{prsty}

\end{document}